# Epistemological Equation for Analysing Uncontrollable States in Complex Systems: Quantifying Cyber Risks from the Internet of Things

Petar Radanliev[1] · David De Roure[1] · Pete Burnap[2] · Omar Santos[3]

## Abstract

The Internet-of-Things (IoT) triggers data protection questions and new types of cyber risks. Cyber risk regulations for the IoT, however, are still in their infancy. This is concerning, because companies integrating IoT devices and services need to perform a self-assessment of its IoT cyber security posture. At present, there are no self-assessment methods for quantifying IoT cyber risk posture. It is considered that IoT represent a complex system with too many uncontrollable risk states for quantitative risk assessment. To enable quantitative risk assessment of uncontrollable risk states in complex and coupled IoT systems, a new epistemological equation is designed and tested though comparative and empirical analysis. The comparative analysis is conducted on national digital strategies, followed by an empirical analysis of cyber risk assessment approaches. The results from the analysis present the current and a target state for IoT systems, followed by a transformation roadmap, describing how IoT systems can achieve the target state with a new epistemological analysis model. The new epistemological analysis approach enables the assessment of uncontrollable risk states in complex IoT systems—which begin to resemble artificial intelligence—and can be used for a quantitative self-assessment of IoT cyber risk posture.

**Keywords** Epistemological analysis · Internet of Things · Risk transformation roadmap · Cyber risk regulations · Empirical analysis · Cyber risk self-assessment · Cyber risk target state

✉ Petar Radanliev
petar.radanliev@oerc.ox.ac.uk

1. Department of Engineering Sciences, Oxford e-Research Centre, University of Oxford, Oxford, England, UK
2. School of Computer Science and Informatics, Cardiff University, Cardiff, Wales, UK
3. Cisco Research Centre, Research Triangle Park, Durham-Raleigh, NC, USA

## 1 Introduction

The Internet-of-Things (IoT) cyber risk is naturally increasing with the increasing digital infrastructure. Cyber risk standardisation and regulation would play a key role in the process of reducing cyber attacks from the IoT. The cyber risk from IoT devices is present across different and sometimes at a higher level in sectors where such risk is unexpected. There is a strong interest in regulating cyber risks and standardising the risk assessments. But cyber risk in complex IoT systems presents too many uncontrollable risk states that are difficult to quantify. By uncontrollable risk, we refer to cyber risk that is still present, even after all cybersecurity recommendations have been followed, and appropriate cyber risk assessment has been conducted on regular basis. An example of the uncontrollable risk state is when a complex and coupled IoT system is exposing the communication network or critical infrastructure, such as compromised IoT device installed in the supply chain—that is out of our control and cybersecurity experts do not have visibility of that IoT device. Risk assessment of uncontrollable risk states is a major challenge and a major limitation of existing literature. Nevertheless, complex IoT systems are already operational in our communication network, and such complex systems are loosely coupled to critical infrastructure (e.g., healthcare during COVID-19), presenting new forms of cyber risk that are often undetected and invisible to security practitioners. To prevent future integration of such complex and coupled IoT systems in the communications network and national critical infrastructure, we need to focus on standardisation of cyber risks, and standardisation of risk assessments (e.g., consider the effect of NIST in the US). Standardisation in this article refers to understanding the best approach for cyber risk assessment in the IoT space. Epistemological analysis is performed of different approaches to determine the best approach for cyber risk assessment in the IoT space. Various IoT systems are being connected to the communications network at a great speed, operating on different communication technologies, creating significant new cyber risks that remain undetected. The lack of a risk assessment approach designed specifically for IoT cyber risks, creates an urgency in understanding what has prevented regulators from creating a standardised approach. There is also a great urgency to understand why existing cyber risk assessment approaches have failed to appropriately risk assess emerging IoT systems.

Hence, the epistemological analysis in this article has two objectives. To identify and capture a target state for cyber risk assessment of complex IoT systems, and to adapt a transformation roadmap for existing cyber risk assessments frameworks, models and standards to include the appropriate risk assessment of various and significantly different IoT systems. For the first objective, to quantify uncontrollable risk states, epistemological equation is presented that decouples risk in complex IoT systems. For the second objective, the epistemological equation risk quantification is discussed and expanded further in the remainder of this article and applied to build a target state for IoT risk assessment and transformation roadmap for advancing existing cyber risk assessment frameworks, models and standards, to include the assessment of cyber risk from complex and coupled IoT systems.

## 2 Review of Related Literature

The research methodology is formulated upon recommendations from existing and related literature on this subject. Since 1989, Woodsmall [1] engaged with the topic of 'Cybernetic Epistemology', discussing the cyber world from epistemological perspective, reviewing the application of control theory and perceiving cybernetics as the theory of feedback systems. In more recent literature, the topic of cybernetics is interlinked with topic of IoT, particularly in the perspective of risk assessment and knowledge management of ethical hacking in a sociotechnical society. To use the words of Abu-Shaqra [2], 'Technoethics' and 'Sensemaking' are required in cyber risk assessment. This author, in his recent Ph.D. thesis, argues for 'the systematization and standardization of an ethical hacking body of knowledge'. Whyte [3] on the other hand argues that there are 'methodological challenges inherent in the multifaceted program' and 'that any answer to this question will vary depending on how one perceives the social science enterprise', resulting with the conclusion that we need to place our research efforts on the 'epistemological and ontological value of different methods lies'. Resnyansky [4] supported this idea for answering the epistemological challenge by designing a conceptual frameworks for social and cultural Big Data analytics. Resnyansky views 'Big Data as an epistemological challenge that stems not only from the sheer volume of digital data but, predominantly, from the proliferation of the narrow-technological and the positivist views on data.' Building upon similar views, Polanyi [5] argues that 'Epistemology is at the heart of the intelligence analysis profession', and proposes a similar conceptual framework, placing 'tacit knowing' and 'personal knowledge' as 'a more precise account for understanding the tacit process of skilfully solving problems of epistemic complexity, along with a deeper appreciation for the personal aspect of knowledge'. Other authors see challenges in these perceptions, for example Daniels [6] analyzes cloaked websites, which are sites published by individuals or groups who conceal authorship to disguise deliberately a hidden political agenda.'. His article titled 'Cloaked websites: propaganda, cyber-racism and epistemology in the digital era', analyses not only how 'cloaked websites conceal a variety of political agendas from a range of perspectives', but also reviews 'cloaked white supremacist sites that disguise cyber-racism' and discusses critical problems related to 'knowledge production and epistemology in the digital era'. The most interesting part of this article is that the article was published more than a decade ago and argues that 'cloaked sites emerge within a social and political context in which it is increasingly difficult to parse fact from propaganda, and this is a particularly pernicious feature when it comes to the cyber-racism of cloaked white supremacist sites'. In the lights of recent events from 2020 to 2021, we can understand that some issues have (unintentionally) been allowed by regulators to develop, grow and evolve. This article is placing focus on why some of these issues have gone undetected by regulators. Why the risk assessment frameworks did not foresee such risks form fake news and simple social engineering attacks by AI driven bots. The focus of this article is on applying the existing knowledge from epistemological techniques, to identify the problems with current cyber risk assessment approaches, and to formulate some of the possible solutions. Although the research fields of epistemology can be seen as distant from the topic of cyber risk assessment, epistemological and ontological analyses are already used in The Royal Canadian Air Force as a cyber warfare schools of thought [7]. This article builds upon these real-world applications of epistemological techniques for improving and bridging the gaps in excising cyber risk assessment approaches.

## 3 Methodology

Current cyber risk concerns are founded on intelligent systems capable of human–computer interactions [8]. The emergence of IoT has resulted with a large number of low-cost connected devices, capable of communicating with each other though a variety of communication protocols (e.g., LoRa, ZigBee, WiFi, LAN, WAN, 5G). Such low-cost IoT devices are often capable of understanding human–computer input, often through remote control (e.g., autonomous vehicles). The low cost has been one of the main triggers for the rapid rise of IoT devices, but the low cost of these intelligent and perceptive devices has also been the main difficulty in protecting these devices from cyber risks. The IoT has created human–computer connection points in modern homes (e.g., smart fridge, smart security cameras), in modern cities (e.g., smart transport), even in the modern wearables (e.g., smart watch). Modern cybersecurity is built with elements of artificial intelligence (AI) such as intrusion detection systems based on user behaviour analysis. Such AI enhanced cybersecurity has proven effective in detecting, responding and preventing both known and unknown cyber threats, but adversaries are increasingly using AI to target connected systems and the low-cost/low-memory nature of the IoT devices makes them increasingly susceptible to cyber attacks.

Further cyber risk assessment challenges emerge from compiling of connected systems, devices and platforms [9]. This creates cyber risk (e.g., from data in transit) [10] and requires standardisation of processes [11]. We combine literature analysis on these topics, with epistemological analysis to uncover the best method to define a risk assessment for IoT cyber risk. Epistemological approach was selected to develop equation for the analysis on the connections between knowledge and information that is based on the aspect of 'truth, belief and justification'. Epistemology is the study of knowledge acquisition, and the three conditions of epistemology are 'truth, belief and justification'. Epistemology is a branch of philosophy that seeks to discover what is known and how it is known. One of the typical epistemological questions is where does knowledge come from? In human perception, knowledge comes from reason and logic, while logic is based on two values: true and false. In the context of the topic (i.e., cyber risk), these two values are absolute and not relative to culture, place, time or persons. In the context of other topics, the true and false topics can be related to epistemic notions and concepts related to people 'beliefs'. Significant epistemological analysis is invested in assessing controversial propositional attitudes emerging from different 'beliefs'. For example, the analysis investigates how we know the right answers in cyber risk assessment, although we do not have sufficient evidence to determine if the real value is 'true' or 'false'. The analysis seeks 'epistemic' justification, in other words, the methodology analyses if 'beliefs' are epistemically justified. This differentiates from the absolutes of 'true' or 'false', because justification of a belief is comparative, i.e., one individual may be justified in believing a certain proposition and another individual may have the same beliefs without any justification.

Many of the cyber risk assessment frameworks have been developed on the basis of expert opinions [12] and are representative of 'truths, beliefs and justifications' that can be relative to culture, place, time or persons. The reason for selecting epistemological approach was to assess the 'truth, belief and justification' of the expert opinions that formulated these frameworks and to determine future trends in existing and new frameworks on cyber risk assessment. There are also new quantitative models emerging on this topic. Many new models are developed without considerations on the (un)availability of probabilistic data. The epistemological approach is selected to analyse the connections between these new models and to determine future trends in quantitative cyber risk assessments.

In the epistemological approach, comparative analysis is applied on the leading digital strategies to identify the cyber risks from IoT systems. The results are integrated into an empirical analysis of cyber risk assessment methods. The aim of the analysis is to identify a new epistemological equation for cyber risk assessment that includes risks from IoT systems. The empirical analysis is conducted with seven cyber risk frameworks and two cyber risk models. The key distinction between frameworks (e.g., NIST) and models (e.g., IoTMM—Internet of Things Micro Mort [13]) is the type of input required and the type of output we can obtain. For example, the majority of cyber risk frameworks at present are applied for qualitative risk assessment. While some of the cyber risk models are also qualitative (e.g., TARA—Threat Assessment and Remediation Analysis [14], OCTAVE—Operationally Critical Threat, Asset, and Vulnerability Evaluation [15]), there are emerging quantitative models for cyber risk assessment/analytics (e.g., FAIR—Factor Analysis of Information Risk Institute approach [16]). The main point of interest for the epistemological analysis are the hybrid cyber risk models (e.g., CVSS—Common Vulnerability Scoring System [17]), that operate almost as calculators for estimating and measuring risk. The hybrid cyber risk models are using qualitative and quantitative methods for risk assessment, and often are designed for assessing specific risks that cannot be assessed by individual approaches. Some of the hybrid models seem to have been built for a specific purpose and for a specific time period. The expectations of the engineers building these hybrid models (at the time of design), is undoubtedly aimed at building a temporary fix, until new data becomes available, that would enable the design of a more comprehensive risk assessment approach. This seems to be the case with the CVSS–hybrid design, which is more than a decade old, with many of its original designers long gone from the project. The epistemological analysis in this article proposes a new approach that would enable the assessment of hybrid models and if their design is up-to-date or requires an update, because new information has become available since its original design. Since the CVSS has become somewhat of a legacy model, that is still used by many cyber risk professionals, and it is still considered by some as the state-of-the-art in risk estimation, the CVSS approach is predominating in the epistemological analysis of this article.

The comparative analysis engages with 15 high-tech national strategies.

While there is novelty in combining methodologies that have not been adapted and integrated, the main novelty of this article is the categorisation and assessment of the connections between knowledge and information that is based on 'truth, belief and justification'. This differentiates from existing quantification models [18], with parameters that are based on expert opinions which can be considered as subjective. In epistemology, there is a prominent distinction between 'subjective knowledge'—which is defined as individual knowledge of subjective states, and 'objective knowledge'—which is defined as knowledge of objective reality. Objective knowledge is designated with the status of knowledge that is supported or proven. While subjective knowledge is considered as unsupported (or weakly supported) knowledge. The objective of the epistemological analysis in this article is to evolve cyber risk assessment form 'subjective knowledge' into the realm of 'objective knowledge'.

## 4 Epistemological Analysis

In the Subsect. 5.5, we conduct a detailed empirical and comparative review of the most prominent cyber risk impact assessment approaches. In this section, for the epistemological analysis, only two cyber risk assessment approaches are selected. The selection was based on their theoretical approach which is best analysed through epistemological analysis as opposed to an empirical analysis. The first approach is the National Institute of Standards and Technology (NIST) framework [19] and the second approach is a vulnerability severity scoring system—the 'Common Vulnerability Scoring System' (CVSS) [17]. Further justification for selecting the two risk assessment methods emerges from their opposing approaches. The NIST approach is a framework that integrates knowledge from most of the cyber risk impact assessment approaches analysed in Subsect. 5.5. The NIST approach is also promoting standardisation of risk assessment, and risk assesses based on the NIST framework are generally acceptable across different US Departments of State. The CVSS is quite an opposite approach. CVSS is industry based and build by private companies and based on the FIRST open training platform. The CVSS is free and open industry-based standard for assessing vulnerabilities and assign scores, allowing the prioritisation of responses and resources. Other major difference between the two approaches is that NIST—National Institute of Standards and Technology's [19], implementation guidance [20] uses a traffic lights system and deliberately stay away from quantifying risk and allocating numerical values to risk. CVSS is a hybrid approach, based on qualitative assessment that derives with a quantitative value associated to risk scores. While we could have included various other cyber risk assessment approaches, the two approaches include most of the strengths and weaknesses that are typically associated with qualitative vs hybrid risk assessment approaches.

While empirical and comparative analysis present some unique advantages for risk assessment (see Sect. 5 for empirical and comparative analysis), the epistemological approach was chosen because of its strengths to analyse the rationality and the justification of beliefs. In this section of the research, we wanted to analyse the cyber risk assessment methods, their validity, the scope, and we wanted to make a clear distinction between justified belief and opinion of the engineers that designed the cyber risk assessment approach. In other words, we wanted to risk assess different cyber risk assessment approaches.

### 4.1 Uncovering a Risk Assessment Method for Uncontrollable States in Complex Systems—Based on Evidentialism and Reliabilism

The epistemological analysis focuses on the knowledge–information connection in regard to nominalising the aspect of 'truth, belief and justification'. The analysis starts with the colour coding in the NIST framework traffic light protocol represent conventional abstractions or in CVSS, a mathematical approximation. In in CVSS, a modified attack vector is allocated to a numerical value of 0.85 for a network metric value, and a numerical value of 0.62 for adjacent network metric value [21]. The question is how was this number determined, and why 0.85 and why 0.62 and why red represents information not for disclosure [22]. These questions emerge, because these units of measurement in effect represent symbols with a defined set of rules in a conventional system, where truths about their validity can be derived from expert opinions, hence proven to be correct. These units of measurement do not, however, represent quantitative units based on statistical methods for predicting uncertainty.

The examples represent a conventional system where symbols are based on true or false, right or wrong. But the rules in the examples describe an ethical system, where the absolute true or false is hard and almost impossible to verify. For example, in the above-described scenario, the number values of 0.85 or 0.62, are not based on probabilistic data. The allocated numbers emerge from an assessment, which is based on multiple statements from experts, but the experts do not claim that these numbers are representative of the attack vectors. It seems more appropriate for the described examples to be perceived as ethical systems, where validity is conditional and output is presented as better or worse, in the form of a colour coded system (e.g., NIST). This is the best result that can be obtained from ethical analysis, where ethical decisions cannot represent sums. Logical proofs are not a valid way of deriving ethical verification. Instead, ethical decisions can be weighted. This can be confirmed by comparing knowledge with understanding. Knowledge in this scenario, refers to the knowledge that red represents information not for disclosure and a numerical value of 0.62 is allocated for adjacent network metric value. Understanding requires that such knowledge can be applied in a meaningful way. But the numerical value of 0.62 does not represent a measurement unit for cyber risk for all adjacent networks. Such numerical value would be case specific and depend on many other factors that the proposed conventional systems are not designed to understand.

This triggers questions to (a) What precisely is 'evidence'? and (b) What amounts to 'evidence' and what does not? In epistemology, the concept of 'evidence' is considered relevant to justified belief and necessary for developing knowledge. In the philosophy of science, 'evidence' is considered to be the means to confirm or disprove scientific theories. The questions analysed in this study are more specifically related to (c) How the cyber security experts distinguish between different types of evidence? In many cases, the empirical evidence is missing, and evidence-based cybersecurity requires rigorous scientific investigations of the effectiveness of cyber risk policies and tools in achieving their intended goals. To guarantee validity of cyber risk evidence, four issues need to be considered in the electronic cybercrime evidence: 'collect strictly according to law, collect electronic evidence comprehensively, invite electronic experts to participate, and ensure the privacy rights of the parties' [23]. Considering that many cyber attacks are unreported and with the case of IoT even undetected, this creates a challenge in applying these established scientific approaches for collecting evidence. Hence, we are taking a broader view of this, starting with how scientists make distinctions regarding what evidence to accept and that which they discard. One approach is the quantitative methods for collecting propositional evidence in explanatory, probabilistic and deductive reasoning. This approach is generally accepted and promoted by the FAIR institute, which looks at the probability of an event occurring, the frequency of the event, the severity of the primary and secondary loss. Other cyber risk assessment approaches that we analysed (e.g., NIST, CVSS, TARA, OCTAVE) consider experience as evidence. While experience can be considered as evidence, such evidence is subjective to the 'Regress argument',[1] where every proposition requires justification, while justification requires support, leading to endless questioning of the evidence, also known as 'Infinite regress'.[2] This triggers even deeper philosophical question, as psychologically we perceive only after we believe. What we expect to see, based on belief, is what we are capable of perceiving. To answer this question, this article engages in epistemological analysis of how the practitioners use a variety of risk assessment approaches to transition from a 'justification of belief' to a 'justification of truth'. To situate the key epistemological challenges facing any practitioner, including the cyber security expert upon this argument, we need to consider that some of the cyber risk assessment models have been designed, tested, verified and adopted in practice in less than a decade. While some of the established risk assessment models have taken many decades to be fully tested and verified prior to being adopted by practitioners. Linking back the discussions to different models and approaches, we apply philosophical analysis to design a set of epistemological equations that would enable cybersecurity experts to adapt existing cyber risk assessment approaches to respond to vulnerabilities of the IoT variety.

For example, knowledge requires 'truth, belief and justification' as individual conditions [24]. Knowledge that a numerical value of 0.62 is 'true' metric value for adjacent network, as the related CVSS approach 'believes', needs to be 'justified' to confirm it does not represent just a guess of luck. Since a numerical value. Justification needs to be based on evidentialism [25, 26], where a proposition e.g., numerical value of 0.62, is epistemically justified as determined entirely by evidence. The debate whether cyber risk standards can be epistemically justified, must be based on the facts and evidence currently available. In evidentialism, epistemic evaluations are separate from moral believes and practical assessments, as epistemically justified evaluations might conflict with moral and practical estimations [25].

---

[1] https://en.wikipedia.org/wiki/Regress_argument.
[2] https://en.wikipedia.org/wiki/Infinite_regress.

**Table 1** Epistemological equation—knowledge/justification of cyber risks

| Knowledge | | | | |
|---|---|---|---|---|
| Understanding | 60 | ↕ | 60 | Understanding |
| Information | 50 | | 50 | Information |
| Quantitative | 40 | | 40 | Qualitative |
| Units of measurement | 30 | | 30 | Symbols |
| Quantitative units | 20 | | 20 | Evidentialism |
| Mathematical model | 10 | | 10 | Reliabilism |
| Justification | | | | |

Analysing the cyber risk assessment approaches with evidentialism is not intended to discard the validity of these standards. Quite the opposite, evidentialism theory includes justified beliefs and experiences as evidence and the prior analysis in this article argues that current cyber risk standards do have knowledge level justification. The argument is that the knowledge level justification of the discussed cyber risk assessment standards, seem to be based on other externalist theories such as reliabilism [27, 28]. This is confirmed with presenting numerical values (e.g., 0.62) represented as metric values, despite the lack of quantitative evidence or other conditions that justify such knowledge. Most cyber security standards follow the reliabilists theory where a cyber risk assessment process can be justified and constitute knowledge, even if the process that makes the assessment reliable is not understood. This is known as the generality problem [28], where a given justification of knowledge can also be identified with different and concurrently operating processes, which may, or may not be statistically reliable. To examine this further, if the statistical model is based on numerical values derived from experts' opinions and qualitative statements, since we cannot verify the validity of these statements, nor the expertise's of the experts, then even if we apply the correct statistical data, we cannot be certain that the result of the analysis represents a justified knowledge. The counter argument to this is, even if we are using a statistically proven model, and we have access to check the formulas, but we apply statistical data that was collected in a manner that cannot be considered as reliable evidence, then we still cannot be certain that the result of the analysis represents a justified knowledge. To make these claims more defensible, we propose to design a new epistemological equation, based on a new process for risk assessment through 'Knowledge' and 'Justification'. This seems necessary, because reliabilism is also associated with the 'the new evil demon problem' [29], where 'we believe ourselves to be doing things that we are not doing'. The discussed cyber risk assessment standards appear to be based on reliabilism, because the standards attribute quantitative knowledge to cyber risk measurements, that would otherwise be considered incapable of measuring quantitatively.

**Table 2** Epistemological equation—truth/belief of cyber risks

| 0.1 | 0.2 | 0.3 | Knowledge justification | 0.4 | 0.5 | 0.6 |
|---|---|---|---|---|---|---|
| Statistical methods | Probabilistic data | Numerical values | | Idea | Expert opinion | Conventional system |
| Truth | ↔ | | | | | Belief |

### 4.2 Epistemological Equation

The differences in these approaches and the conclusions that are derived from the epistemological equation are categorised and numbered in Tables 1 and 2 to build a four-quadrant graph for the comparative and empirical analysis of digital strategies and cyber risk assessment approaches. Table 1 presents the described processes for risk assessment through 'Knowledge' and 'Justification'.

Table 2 categorises, numbers and presents the different processes described in the epistemological analysis, for determining between 'Truth' and 'Belief'.

The described categorisations emerge from the epistemological analysis, but the numbering from 0.1 to 0.6 and from 10 to 60 represents the simplest form of data input for designing a four-quadrant graph. We do not associate any value to the numbering, we are simply using the four-quadrant graph method to populate the fields for conducting and driving with meanings from the comparative empirical study. This concludes epistemological analysis and presents the background for the comparative empirical study, which is the bedrock of the paper as a whole.

## 5 Comparative Empirical Analysis of IoT Cyber Risk Assessments in Digital Strategies

This section apples the 'Knowledge' and 'Justification' four-quadrant graph with comparative analysis of IoT cyber risk in digital strategies and empirical analysis of cyber risk impact assessment approaches. The analysis is focused on identifying evidentialism and reliabilism in cyber risk assessments of digital strategies.

### 5.1 Comparative Analysis

The digital strategies included in the comparative analysis are selected based on their representation as comprehensive and well-documented initiatives. The research identified additional digital strategies, but the lack of details provided by subsequent national governments limits their potential in a comparative empirical study. To avoid repetition, in the comparative analysis below existing abbreviations are used from existing literature [30] and include: New National Technology Initiative (NTI) [31]; New Industrial Revolution (NIR) [32]; Made in China 2025 [33]; Fabbrica

Intelligente [34]; Industrie Conectada 4.0 [35]; Made Different [36]; Factories of the Future 4.0 [37]; New France Industrial (NFI) [38]; Industrial Value Chain Initiative [39]; New Robot Strategy (NRS) [40] and RRI [41]; Digital Catapults [42]; Digital Strategy [43]; Made Smarter [44]; Industrie 4.0 [45]; Internet Consortium (IIC, 2017); and Advanced Manufacturing Partnership [46].

### 5.2 Divergence in Cyber Risk Regulations

First, the comparative analysis identified a divergence in different approaches. The Russian NTI strategy diverges mostly from the other digital strategies. The NTI focuses on market creation instead of technology creation and promotes regulations vs loosely defined standards. In addition, the NTI does not assess risks in real-time cloud networks. In the IIC, Industrie 4.0, and DCMS, cyber security strategies are proposed in the form of cloud-computing platforms, but there is no mention on how these cloud computing platforms actually provide/allow for cyber risk impact assessment.

Second, some digital strategies have explicitly developed cyber risk architectures, e.g., IIC and impact assessment e.g., Industrie 4.0. Other focus on loosely defined risk assessment that emerge from forums, such as in the case of IVI; or blogs, in the case of Made Different; or surveys in digital catapults. Some strategies promote cyber risk assessment through workgroups (e.g., IVI), which directs qualitative risk assessment. While other promote activities in the format of testbeds, (e.g., IIC or digital catapults) that enable quantitative risk assessment. The direction of risk assessment is decided by the assessment activities, e.g., workgroups vs testbeds, or economic impact vs risk assessing key projects in the digital industry, e.g., Fabbrica Intelligente. The different approaches to risk assessment are a result of the differences in focus. The IIC focuses on core IoT industries; while NFI, the digital catapults and the NTI, all focus on the development of key IoT technologies. Made in China 2025 focuses on tech sectors, while the made different focuses on key IoT transformations. The digital catapults and NFI promote economic risk impact assessments. While the NTI promotes market risk assessment. Finally, some digital strategies are very narrowly focused on futuristic new technologies (e.g., NRS and RRI) that do not yet exist. Hence, such risks cannot be regulated, because we can only attempt to forecast the expected cyber risks.

### 5.3 Lack of Data Strategies for Collecting Probabilistic Data

This divergence is cyber risk regulations in IoT strategies leads to private sector and national statistical agencies being unable to develop the required data strategies that would enable quantitative risk assessment. The lack of appropriate data strategies leads to digital strategies lacking documentation on probabilistic data and the risk assessment appear disorganised. Such arguments are also present in literature, stating that in the present state, digital strategies appear unprepared for assessing new cyber risks [47]. As the first step in the pursuit of existing regulations for creating data strategies, we referred to the NTI recommendation for creating a direct electronic open feedback for changing data strategies. But prior to analysing any feedback, the main IoT risk elements of each IoT strategy need to be compounded into categories representing the most prominent IoT cyber risks. However, the compelling of data into these categories is quite challenging, as some strategies, represent a collection of descriptive explanations and do not provide explicit IoT cyber risks.

**Table 3** Digital strategies assessed and categorised with the epistemological equation on 'Knowledge' and 'Justification' and 'Truth' and 'Belief'

| | Digital strategies | Cyber risk assessments | |
| --- | --- | --- | --- |
| | | 'Knowledge' and 'Justification' | 'Truth' and 'Belief' |
| Germany | Industrie 4.0 | 10 | 0.1 |
| USA | Industrial Internet Consortium | 30 | 0.2 |
| | Advanced Manufacturing Partnership | 40 | 0.4 |
| UK | Digital Catapults | 10 | 0.1 |
| | UK Digital Strategy | 20 | 0.6 |
| Japan | Industrial Value Chain Initiative | 20 | 0.6 |
| | New Robot Strategy and RRI | 10 | 0.1 |
| France | New France Industrial—NFI | 20 | 0.4 |
| Nederland | Factories of the Future 4.0 | 10 | 0.1 |
| Belgium | Made Different | 20 | 0.5 |
| Spain | Industrie Conectada 4.0 | 50 | 0.4 |
| Italy | Fabbrica Intelligente | 40 | 0.6 |
| China | Made in China 2025 | 10 | 0.1 |
| G20 | Industrial Revolution | 60 | 0.4 |
| Russia | National Technology Initiative—NTI | 30 | 0.6 |

To resolve this issue, we compare the most prominent cyber risk assessment method, and we use the findings as reference points to define a standardisation or sometimes contrasting assessments of cyber risk from IoT systems.

### 5.4 Populating the Epistemological Equation with R-Squared Values

The findings from the comparative analysis are populated in a four-quadrant graph and the risk assessment approaches are assessed and categorised with the 'Knowledge' and 'Justification' (Table 1) and 'Truth' and 'Belief' (Table 2) epistemological equation.

Table 3 presents the risk assessments approaches epistemological categorisations emerging from the comparative analysis of the digital strategies. The risk assessments categorisations in Table 3 are directly related with the 'Knowledge' and 'Justification' and 'Truth' and 'Belief' of with the epistemological equation in Tables 1

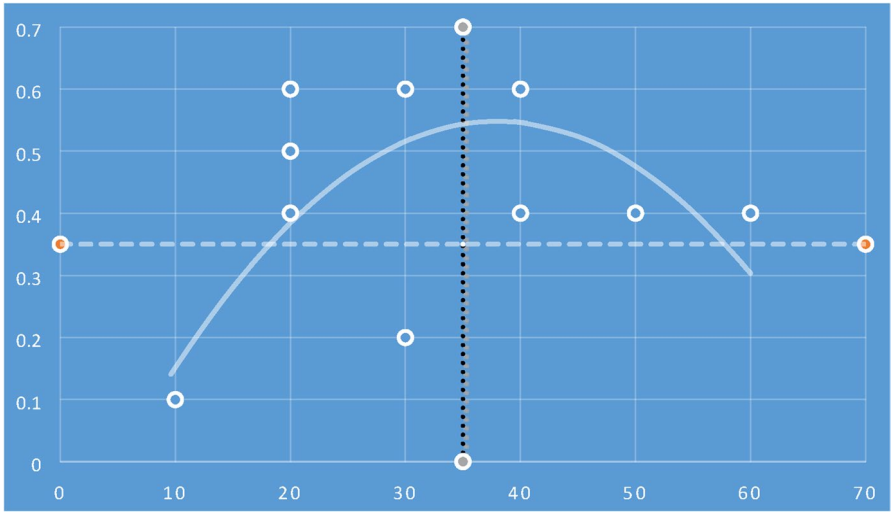

**Fig. 1** Four-quadrant graph displaying the R-squared values of risk assessment approaches in digital strategies—analysed with the 'Knowledge' and 'Justification' and 'Truth' and 'Belief' epistemological equation

and 2. The epistemological data from Table 3 is populated in a four-quadrant graph displaying a polynomial curve of the R-squared values (Fig. 1).

The data points in Fig. 1 four-quadrant graph are extracted from Table 3 and the R-squared values presents a clear trend in qualitative and quantitative methods towards risk assessments based on knowledge understanding and less focus on justification of truth and belief. The epistemological equation shows a clear trend of lower representation in the justification of truth (quantitative and mathematical models) and in the justification of belief (evidentialism and reliabilism). The next step conducts empirical analysis of cyber risk impact assessment approaches, to identify if the R-squared values of the epistemological equation are caused by and representative of the current state of the art in cyber risk assessment methods.

### 5.5 Empirical Analysis of Cyber Risk Impact Assessment Approaches

The empirical analysis initiates with identifying cyber security frameworks and comparing with most recent cyber security literature on this subject. The empirical analysis includes qualitative and quantitative approaches to measuring cyber risk. Some of the analysed frameworks propose diverse qualitative methods, such as OCTAVE—measures cyber risk through workshops. The TARA methodology applies a threat matrix. The CVSS applies expert's opinions, presented as statements, where each statement is allocated a level of cyber risk and the calculator assesses the overall level of risk form all statements. Considering the lack of more precise methods, at present such approaches represent the state of the art for IoT cyber risk assessment. The state of the art in current risk estimation is based on the high, medium, low scales (also known as the traffic lights system or colour system). The above analysis suggests that there is a disconnection from the aforementioned risk assessment approaches and the risk assessment requirements for justification of truth (quantitative and mathematical models) and the justification of belief (evidentialism and reliabilism). However, the analysis provides an explanation for the trend in digital strategies towards risk assessments based on knowledge understanding and less focus on justification of truth and belief through quantitative and mathematical models or evidentialism and reliabilism. Table 4 presents the epistemological categorisations emerging from the empirical analysis of the risk assessments approaches.

Table 4 Cyber risk assessment approaches assessed and categorised with the epistemological equation on 'Knowledge' and 'Justification' and 'Truth' and 'Belief'

| Cyber risk method | Knowledge–justification | Truth–belief |
| --- | --- | --- |
| FAIR | 20 | 0.4 |
| CMMI | 30 | 0.5 |
| CVSS | 20 | 0.6 |
| ISO | 40 | 0.4 |
| NIST | 60 | 0.5 |
| OCTAVE | 50 | 0.4 |
| TARA | 10 | 0.4 |
| RiskLens | 10 | 0.3 |
| CyVaR | 10 | 0.1 |

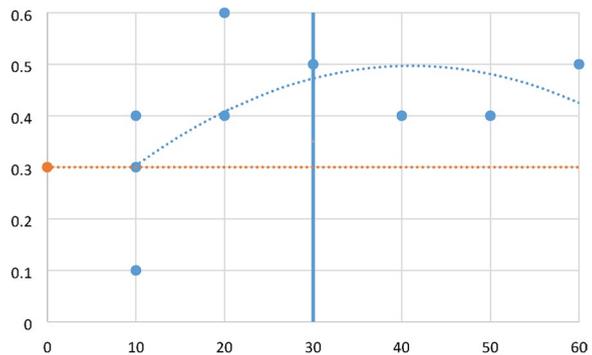

Fig. 2 Four-quadrant graph displaying the R-squared values of cyber risk assessment methods—analysed with the 'Knowledge' and 'Justification' and 'Truth' and 'Belief' epistemological equation

The epistemological categorisations of risk assessments approaches (in Table 4) presents the categorisations of the empirical analysis that are directly related with the epistemological equation in Tables 1 and 2. The epistemological data from Table 4 are populated in a four-quadrant graph in Fig. 2 displaying the R-squared values to compare the results with the R-squared values of the four-quadrant graph in Fig. 1.

The data points in Fig. 2 four-quadrant graph are extracted from Table 4 and presents the polynomial curve of the R-squared values of the epistemological equation. Similar to Table 1, the polynomial curve confirms that of the R-squared values of risk assessment methods are based on knowledge understanding and not on justification of truth and belief. The polynomial curve in Table 2 shows much lower representation in the justification of truth (quantitative and mathematical models) and not a single method in the quadrant for justification of belief (evidentialism and reliabilism). The empirical analysis shows that qualitative approaches present a lack of precision, e.g., one expert perception of a threat as low might not conform to another expert perception or belief. In other words, Fig. 2 can be interpreted as a confirmation that qualitative methods are promoting risk assessments that can only be verified with quantitative data. Since there is not a single qualitative method that can be verified with the epistemological equation as representing a justification of belief (evidentialism and reliabilism), then the question is whether these methods can be verified with justification of truth (quantitative and mathematical models).

### 5.6 Target State—Cyber Risk Assessment that Includes IoT Risk Based on Justification of Truth

The difficulty in verifying these methods with justification of truth (quantitative and mathematical models) is that mathematical models lack sufficient probabilistic data and given the complexity of the analysis in interconnected systems, there are very few mathematical models that claim success in quantitative risk estimation and analysis. However, if such interconnected verification was possible, the process of combining the strengths of the present approaches would present an improved cyber risk assessment that also includes IoT risk. For example, the risk assessment approaches that are based on justification of truth, do not calculate the cyber risk from shared infrastructure, e.g., supply chains. The Exostar system [48] can be used for complimenting these approaches and covering the supply chain aspect of cyber risk. The overall current state of cyber maturity can be verified with the CMMI—Capability Maturity Model Integrated [49], which integrates five levels of the original CMM—Capability Maturity Model [50]. To reach the required cyber security maturity level, the current cyber state can be transformed into a given a target cyber state by applying the NIST implementation guidance [20].

### 5.7 Target State of Cyber Risk Assessment Based on the Epistemological Equation

To devise a target state of cyber risk assessment based on justification of truth, we refer to quantitative cyber risk assessment models, starting with the IoTMM for quantitative IoT cyber risk assessment and the FAIR model for a quantitative overall cyber risk assessment (FAIR uses the RiskLens [51], and CyVaR—Cyber VaR [52] models). The IoTMM and FAIR models are complementary to the work of NIST and ISO. For example, ISO 27001 and ISO 27032—International Organisation for Standardisation [53]. The ISO 27032 provides specific cyber risk recommendations and ISO 27001 sets requirements for cyber security, but only FAIR provides recommendations for quantitative cyber risk estimation, and only the IoTMM provides a free and publicly available model for IoT cyber risk assessment.

The following section devises a target state of cyber risk assessment based on the epistemological equation and presents a transformational roadmap for integrating knowledge and understanding from frameworks and models that have focused justification of belief, but failed to achieve that with evidentialism and reliabilism. The transformation roadmap is based on integrating the aforementioned risk assessment approaches in the justification of truth quadrant of the epistemological equation.

### 5.8 Transformation Roadmap for Standardisation of IoT Risk Impact Assessment

The transformation roadmap for changing from current to a target state of cyber risk assessment, is presented though strengths, weaknesses, opportunities and threats (SWOT) analysis. The IoTMM model is not included in this stage of the analysis, because the IoTMM already assesses IoT risk as justification of truth in a standalone model. To enable the transformation from justification of belief to justification of truth, first the core cyber impact assessment concepts are related to risk assessment areas based on quantitative and mathematical methods for justification of truth. Then, the transformation roadmap is designed through SWOT analysis. The transformation areas for justification of truth are defined as how to: identify, manage, estimate, and prioritise IoT cyber risks. These areas are described as:

- Risk identification (measure)—current state of IoT cyber risk.
- Risk management (standardise)—target state for IoT cyber risk assessment approach.
- Risk estimation (compute)—quantify IoT cyber risk in the target state.
- Risk prioritisation (strategy)—transform IoT risk from current state into a target state.

Tables 5, 6, 7, 8 and 9 address two objectives of this paper. First, it presents a target state of cyber risk assessment, based on integrating the aforementioned risk assessment approaches and the requirements found in digital strategies. Second, the transformation roadmap presents a new approach for reaching the target state in IoT risk assessment. The transformational roadmap also derives with a new approach for analysing the strengths and weaknesses of existing cyber risk assessment approaches for identifying, managing, estimating, and prioritising IoT risks.

## 6 Discussion on Results with Bibliometric Analysis

To analyse our findings with similar research in this field, we conducted bibliometric data mining on the Web of Science Core Collection. Our first search included the topics of epistemology and cyber risk and resulted with 0 records. Our second search included the topics epistemology and cybersecurity and resulted with only four records, none of the records was related to risk assessment. We continued with changing the search parameters to: epistemology and risk maturity, which also resulted with 0 records; epistemology and internet of things, and finally, with epistemology and IoT, which resulted with 6 and 3 records. These data records were too few for bibliometric analysis, and none of the records was related to risk assessment

**Table 5** Target state for IoT cyber risk assessment based on the epistemological equation

| Target state | | | | |
|---|---|---|---|---|
| Vectors | Vector 1 | Vector 2 | Vector 3 | Vector 4 |
| | Risk identification | Risk management | Risk estimation | Risk prioritisation |
| | Measure | Standardise | Compute | Strategy |
| Risk models | | | | |
| OCTAVE | Asset-based threat profiles | N/A | Qualitative | N/A |
| TARA | Cyber threat susceptibility assessment (CTSA) | Cyber risk remediation analysis (CRRA) | (a) Template threats (b) Scoring system (c) Threat matrix | Mission assurance engineering strategies (MAE) |
| CVSS | Base metrics | Mathematical approximation | Qualitative | N/A |
| Exostar | Managed access gateway (MAG) | Partner information management (PIM) and vendor quality management (VQM) | N/A | Source-to-pay software-as-a-service solution |
| CMMI and CMM | Maturity models | ISO 15504—SPICE | Maturity levels | N/A |
| NIST | Categorising | Assembling | Compliance | Compliance |
| FAIR | Financial | Compliance | Quantitative | Level of exposure |
| RiskLens | Probabilistic data | N/A | Monte Carlo simulations | N/A |
| CyVaR | Probabilistic data | Value at risk model | Monte Carlo simulations | N/A |
| ISO | ISO 27032 | ISO 27001 | Compliance | ISO 27031 |
| IoTMM | Probabilistic data | Value at risk model | Monte Carlo simulations | Micro Mort model |

**Table 5** (continued)

| Target state | | | | |
|---|---|---|---|---|
| Vectors | Vector 1 | Vector 2 | Vector 3 | Vector 4 |
| | Risk identification | Risk management | Risk estimation | Risk prioritisation |
| | Measure | Standardise | Compute | Strategy |

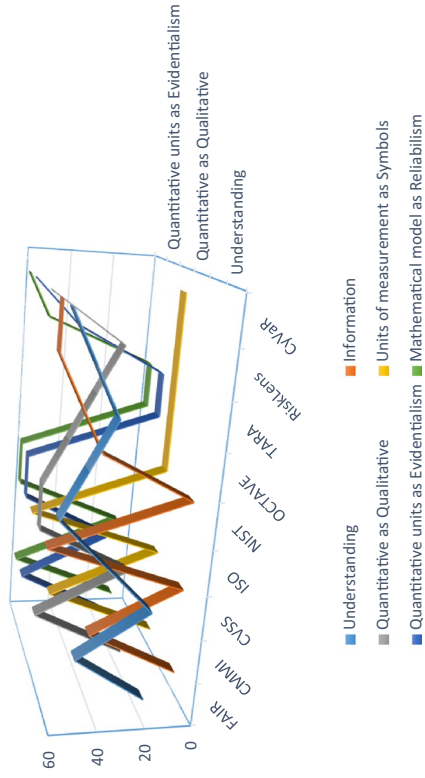

**3D line of the target state based on the epistemological equation**

**Table 6** Transformation roadmap for reaching the target state for justification of truth in IoT cyber risk assessment—reaching the target state for IoT cyber risk assessment through the implementation tiers

Transformational roadmap for IoT risk assessment

Implementation tiers—strengths for justification of truth:

OCTAVE has developed a standardised questionnaire that can be applied to investigate and categorise IoT risk impact areas

TARA is a predictive framework that enables targeting of the most crucial IoT exposures, as opposed to promoting the defence of all possible vulnerabilities

CVSS can be used to translate qualitative input into a numerical score reflecting severity and characteristics of IoT vulnerabilities

Exostar system can be used to assess, measure, and mitigate IoT risk in real-time across multi-tier partner and supplier networks and to determine the gaps between cybersecurity posture and regulatory compliance

CMMI can be used to simultaneously assess the full IoT product development life cycle risk and to measure multiple as opposed to stand-alone improvements

The NIST framework can be used in assessing IoT cyber risk, but more valuable in managing IoT cyber risks

FAIR model promotes a quantitative, risk based, acceptable level of loss exposure that can be adopted for IoT risk

ISO can be used to promote standardisation of IoT cyber risk and to reflect on international experience and knowledge

RiskLense presents a quantitative assessment with Monte Carlo simulations and can be adopted for IoT risk

CyVaR presents a method to quantitatively assess risk with Monte Carlo simulations and can be adopted for IoT risk

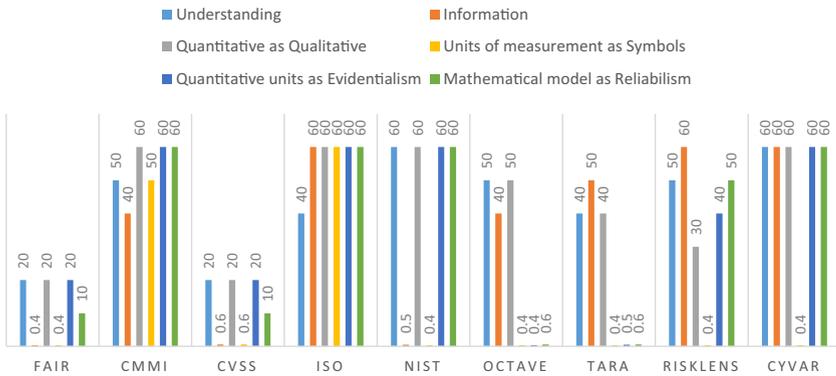

or cybersecurity. We found 68 records on epistemology and risk assessment, but no records related to health, ethics, medicine, and only one record related to epistemology and risk assessment in information technology. We conducted numerous searches on the topics covered in this article, but we only found limited records on (1) epistemology and information technology; (2) epistemology and complex

Table 7 Implementation tiers—weaknesses in current approaches for cyber risk assessment

Implementation tiers—weaknesses for justification of truth

OCTAVE fails to provide a quantification method for calculating cyber risks—including IoT risk

TARA fails to quantify the impact of cyber risks—including IoT risk

CVSS contains scoring range between 0.0 and 10.0, but is based on a 3-level system and because the score is derived from a limited number of variables, it creates dissimilar vulnerabilities receiving similar score

Exostar system does not assess enterprises own cyber risk exposure. Instead, it helps enterprises to manage risk by understanding the strengths and vulnerabilities of their supply chain partners

CMMI does not explain how to implement improvements, but only indicates where improvements are needed. The improvements are not methodological processes and the actual processes an enterprise chooses depend on multiple factors. The CMMI simply does not map the IoT risk assessment processes

NIST framework is documented, not an automated tool and does not contain an impact assessment model for quantifying IoT cyber risk

FAIR framework promotes standardisation of quantitative models, but is difficult to use for IoT risk assessment, because it is not as documented as other frameworks

ISO is based on voluntary shared knowledge and is consensus based. International standardisation of IoT risk assessment requires a level of compulsory compliance

RiskLense contains a lack of details on the algorithm supporting its risk assessment. Process for IoT risk assessment is not included

CyVaR has the potential issue of a lack of the required IoT risk data to perform adequate and comprehensive assessments

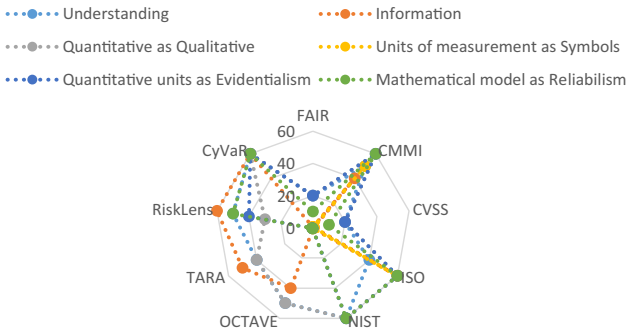

systems—385 records; and (3) epistemology and computer science—165 records. Hence, we used the data records on epistemology and information technology, which resulted with 285 records, for bibliometric analysis with R Studio [55]. Then, we changed the bibliometric program to VOSviewer [56] to analyse the data records on epistemology, complex systems and computer science. To analyse all 23,608 data records on epistemology, we used the Web of Science analyse results data mining tool (Fig. 3).

From the bibliometric analysis with the Web of Science analyse results data mining tool in Fig. 3, we can see the epistemology has been used in computer science and information science research, but not as much in cyber risk and security. Therefore, our article contributes to this area of knowledge, by applying epistemology in cyber risk and security research. We continued the bibliometric analysis with the data records on epistemology and information technology, with R Studio Fig. 4.

Table 8  Implementation tiers—opportunities in current approaches for cyber risk assessment

| Implementation tiers—opportunities for justification of truth |
| --- |
| OCTAVE is free and can be used as the foundation risk-assessment component or process for IoT risk assessment |
| TARA can be implemented as a complementary method IoT risk assessment, in combination with OCTAVE |
| CVSS currently has a 3-level scoring system, and as such the biggest opportunity is to integrate IoT risk in the form of more levels in the calculator to represent cyber risk with greater precision |
| Exostar system could evolve into a system that assesses enterprises own IoT cyber risk exposure, while enabling the assessment of cyber risk from supply chain partners |
| CMMI is related to ISO 9001. The ISO 9001 specifies a minimal acceptable quality level, while CMMI specifies continuous process improvement. Biggest opportunity is to adapt CMMI with continuous updates from ISO 9001 and with emerging IoT standards |
| The NIST is based on an extensive use of acronyms, which can be confusing and require a detailed understanding of the standards referred to in the acronyms. Hence, the greatest opportunity would be adding IoT risk acronyms in the process of simplifying the design. This could be done by replacing the acronyms with a new user-friendly tool to incorporate a fully automated guidance process (e.g., such as CVSS calculator) |
| FAIR is complementary to existing risk frameworks and applies knowledge from existing quantitative models. This represents an opportunity for developing a standardisation IoT risk reference architecture |
| ISO could evolve into an international standardisation of IoT cyber risk/security framework |
| RiskLense could evolve into the first standardised quantitative model for IoT cyber risk assessment. More academic research is required on this model to define and disclose the algorithm. This would increase the acceptance of this model, as academic research would enable the model to be verified and validated |
| CyVaR needs to be adapted and modified to include units of measurement for IoT cyber risk vectors |

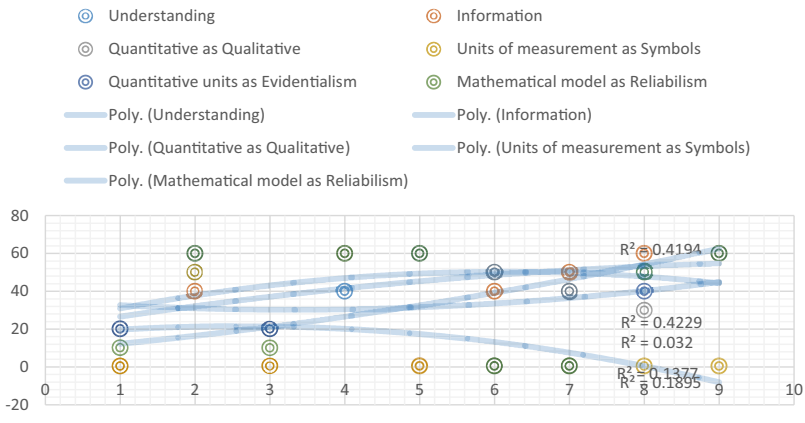

**Table 9** Implementation tiers—threats for justification of truth in current approaches for cyber risk assessment

| Implementation tiers—threats for justification of truth |
|---|
| OCTAVE method is complex and takes time to understand. This is the main weakness as it is a qualitative method that does not provide mathematical or financial modelling |
| TARA focuses on reducing cost by covering only the exposures that are most likely to occur, but the assessment ignores IoT risks |
| CVSS converting qualitative data into a quantitative result, with relatively low-level mathematical approximation, could create a false level of security |
| Exostar system uses third-party sources to provide insights in the cyber health and viability of supply chain partners. The validity of the data depends on the third-party sources and if this cyber data is incomplete or compromised, the insights would also be compromised |
| CMMI measures are easy to recognise but difficult to develop. For instance, CMMI does not provide guidance on how to implement improvements, it simply indicates where improvements are required |
| NIST as a documented model, depends on many documents being continuously updated. Unless it evolves into a more automated process, the framework would need constantly to be reviewed and updated as new technology and laws emerge |
| FAIR depends on a computational engine for calculating risk and a model for analysing complex risk scenarios RiskLens [51]. RiskLens [51] is a commercial product and the software comes at a cost. Standardisation of commercial products could create disadvantages for small enterprises that lack resources of large enterprises. Small enterprises may choose free models such as OCTAVE |
| ISO contains members from 161 countries and 778 technical committees and subcommittees. This presents a major challenge in coordination and integration of specific standards [54] |
| RiskLense, without the academic peer-review rigour and industry expert review, represents a model that is very difficult to verify and validate. Without such validation, the results would be questionable |
| CyVaR is a fairly complicated approach and unless simplified, in a software format, similar to the CVSS, it could be difficult to implement as a standard model for cyber impact risk assessment |

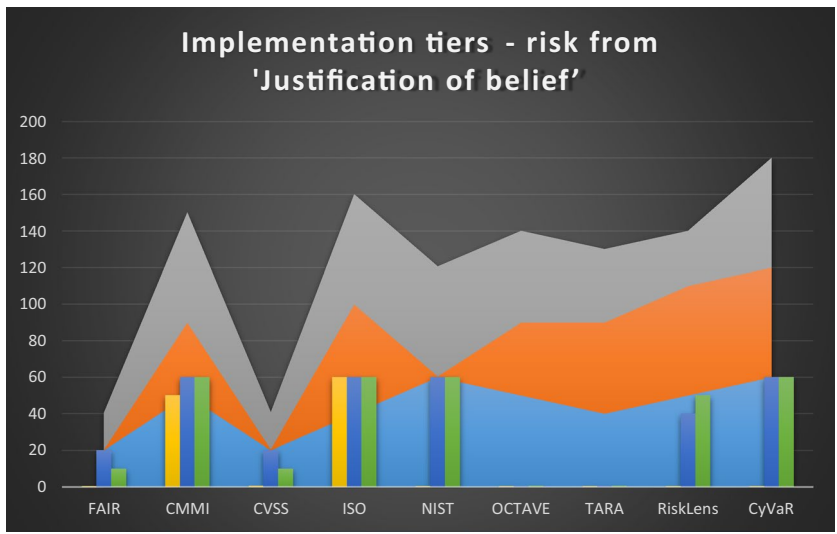

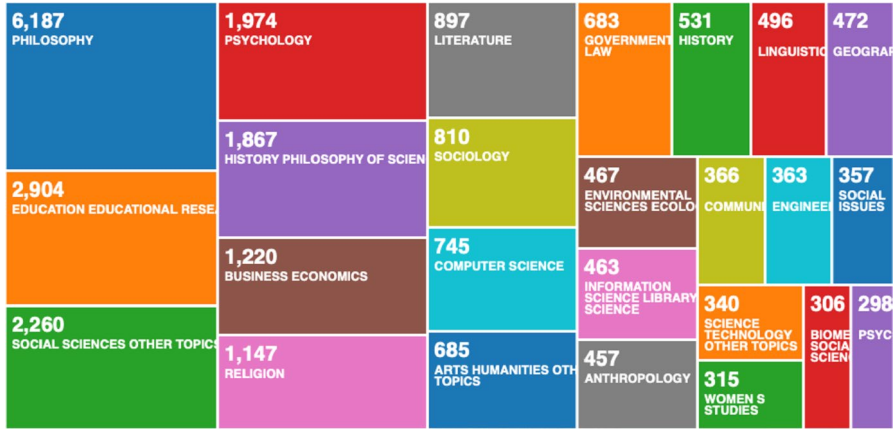

**Fig. 3** Web of Science analyse results data mining tool—discussion on article results with bibliometric analysis

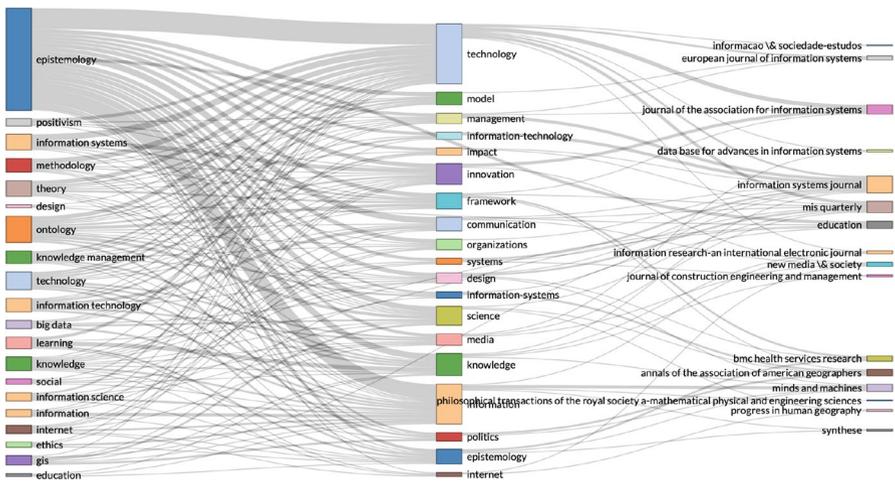

**Fig. 4** Three-fields plot: bibliometric analysis with R Studio on epistemology and information technology data records

In the three-field plot (Fig. 4), we wanted to identify from the data records, the topics related on epistemology and information technology. We separated the statistical analysis by keywords extracted from the text in the data records—on the left, keywords listed in the articles—in the middle, and the journals that published most articles on these topics. What we can see in Fig. 4, is that similar research has been conducted on a variety of topics, but not as much in cyber risk and security. This strengthened our argument that this article contributes to a gap in knowledge, by applying epistemology in cyber risk and security research. We illustrate this further in Fig. 5, by separating the data records in topic dendrogram.

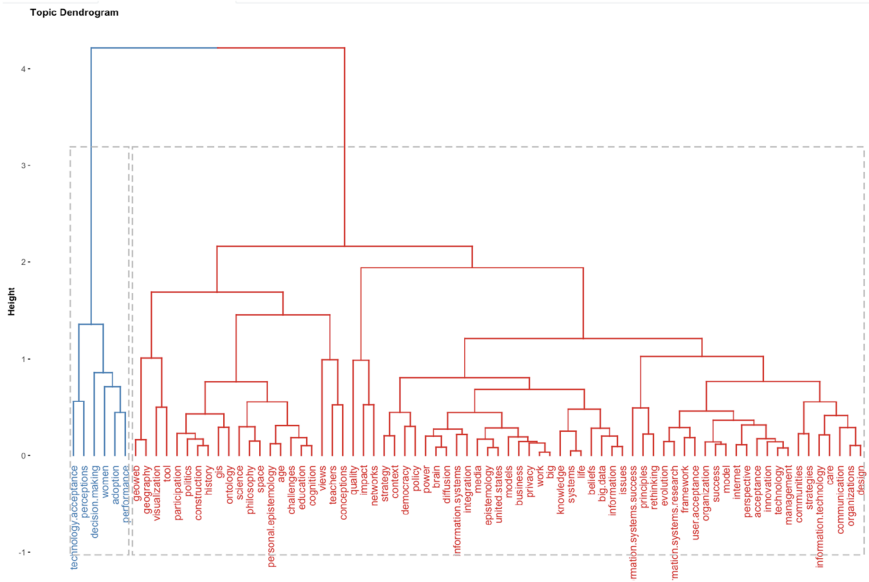

**Fig. 5** Topic dendrogram: factorial analysis with R Studio on epistemology and information technology data records

The topic dendrogram is constructed with factorial analysis, and the visualisation presents categories of similar research, on a variety of topics, except cyber risk and security. We advance the discussion on results though bibliometric analysis, with the VOSviewer computer software. We used the VOSviewer to analyse in combination the data records on epistemology and complex systems—385 records, and epistemology and computer science—165 records. In the two images from VOSviewer in Fig. 6, we analysed the authors stated keywords from the combined data records.

With VOSviewer, in Fig. 6, we could have analysed all keywords that appear in the text of the data records, but we did not want to identify the individual representativeness of these topics, we wanted to identify if these topics appear in the data records. If we found strong representation, we could have analysed the data records further, but we could not identify any representation of the topics on cyber risk and security, in the data records on epistemology, complex systems and computer science. These bibliometric visualisations, reemphasise our rationale and our argument that this research, covers a gap in current knowledge.

## 7 Discussion on the Limitations of Statistical Analysis to Generate 'Truths' in Relation to Risk Assessments

The epistemological analysis and the comparative empirical analysis in this article made some interesting discoveries that seem to benefit the quantitative approaches for cyber risk assessment. In reality, however, the most reliable risk assessments are based on quantitative and qualitative approaches. While quantitative assessment presents multiple advantages, we cannot guarantee with absolute certainty that the evidence is correct and the quantitative formulas from 'black box' models are correct. The increased use of AI in cyber risk assessment necessitates a new discussion on how AI can be used by adversaries to prevent our cybersecurity. If adversaries can access the AI training data, or the algorithm parameters, they can easily tamper the data to prevent detection by AI enhanced cybersecurity. Considering that most AI-based cybersecurity is using identical training data, from legally approved data collection sources, and this training data is often publicly available in open access, then we can easily imagine how adversaries will use this to their advantages.

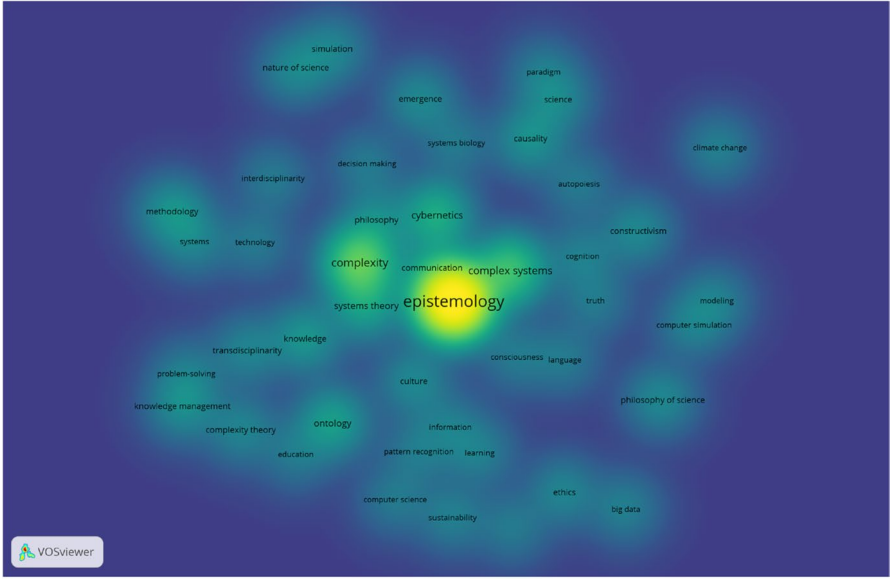

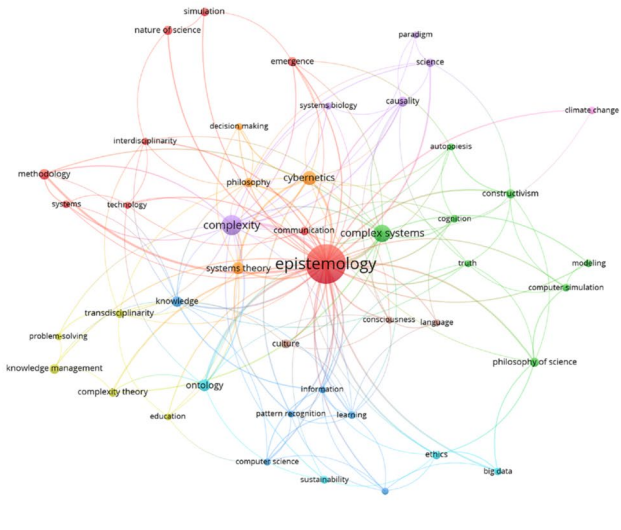

**Fig. 6** VOSviewer bibliometric analysis of data records on epistemology, complex systems and computer science

Within security analysis, there are many occasions when there have been assessments based on an abundance of data, which on the surface would allow for a more reliable assessment (compared to assessments based on insufficient data). Yet the notion that an abundance of data is necessarily superior or better for making accurate assessments is not always the case. If we consider the assessment of WMD in Iraq—there was lots of data to paint the picture that Saddam Hussein was hiding weapon facilities. The 'data' were strong, yet the assessments as the various government inquiries revealed were 'dead wrong'. In contrast, there was very little data indicating Osama Bin Laden was in a compound in Abbottabad—yet he was found to be there.

Section 4 engages into a deep dive examination of the epistemological analysis of cyber risk assessment and the comparative empirical analysis in Sect. 5 challenges the perception of statistical analysis as a 'gold standard', to build upon the epistemological equations, with respect to advancing current cyber risk assessment with a model for determining 'justified true belief' risk assessments. There are some award-winning books, based on similar research studies, but with a more specific quantitative topics e.g., Black Swan events [57]; Superforecasting [58]. These sub-topics centre around the fundamental issues of the value and limitation of statistical analysis regarding security threats, including those threats of a 'cyber' variety.

Related to this discussion, we examine two case studies, focused on different types of models which underpin risk assessments. The examination is centered around the effectiveness of the applied cyber risk assessment models to predict 'correct' or 'false', in the field of IoT cyber security. The first case study is with Amazon supply chain tracking and traceability with IoT-enabled blockchain on AWS.[3] We conducted a detailed search on the AWS web pages, and on Google, using a variety of different keywords—and we failed to identify what risk assessment approach has been used by AWS in their IoT-enabled blockchain. The only information that kept appearing online, was that AWS partnered with Deloitte to ensure safety and security of the project. We conducted a detailed search on the Deloitte web site, and we found no records on the type of risk assessment performed, or whether a risk assessment was performed at all. This is a clear example of a 'black box' risk assessment approach, with no details on how evidence has been collected or analysed.

For the second case study we exanimated, we did not search for an IoT project, instead, we searched specifically for a IoT project that included to IoT risk assessment. We identified a case study of a risk assessment in IoT of a collaborative robot system from the ECLIPSE SAM Virtual Conference.[4] The case study risk assessed

---

[3] https://aws.amazon.com/blogs/apn/supply-chain-tracking-and-traceability-with-iot-enabled-blockchain-on-aws/.

[4] https://www.slideshare.net/BrainIoT/samiot-risk-assessment-in-iot-case-study-collaborative-robots-system.

IoT devices operating on multiple communication technologies (e.g., LoRa, NFC), using a variety of security state-of-the-art common security standards (e.g., ISO/IEC 27002, NIST SP 800-30/82, and specific IoT security standards (e.g., ISO/IEC 30128). The collaborative robot system also used seven different risk assessment methods (including OCTAVE) and concluded that the methods are too generic, and they fail to anticipate the complexities of IoT systems. The project resulted with a new design of a more specific risk assessment method, specific for IoT assets. This outcome is not surprising, considering the results of our analysis, which confirmed that even the most specialised hybrid risk assessment models, are based on 'truth, belief and justification' that seems relative to culture, place, time or persons designing the approach.

Building upon the case study discussion, we need to clarify the unique challenges of cybersecurity in IoT systems. The reality is that IoT devices can be loosely coupled to perform a certain task and have the integrated capability to break that connection when the task is complete. This level of temporality has various implications for network security and combined with the low-cost/low-memory nature of IoT devices, makes the security of IoT systems particularly challenging. In addition, the dynamism of IoT systems/devices presents another challenge to the cyber security practitioner. For risk assessments to be effective, they would need to, where possible, consider and predict (i.e., forecast) the likelihood of certain new IoT systems emerging within the cyber security ecosystem. Predicting new and emerging cyber risks from IoT systems becomes of even greater importance when we consider that cyber risks from some IoT systems remain undetected by cybersecurity professionals. For example, from our discussions with cybersecurity experts, we discovered that in most instances, cybersecurity experts are not even notified when a new IoT device is installed, because we do not have regulations and policies in place to make reporting such installations compulsory.

Although this article cannot provide solutions to all issues related to cyber risk assessment of IoT systems, we make some exceptional observations and arguments regarding a roadmap for transitioning to a target state cyber security threat assessment.

## 8 Conclusion

Current literature contains significant limitations on risk assessment of complex and coupled IoT systems. The challenges faced in adapting existing cyber risk frameworks, models and standards for assessing IoT risk, are predominated by complexities from the IoT abilities to connect through various different communication technologies and protocols. In addition to various other complexities, the low-cost/low-memory nature of IoT devices creates significant challenges in securing such high-tech devices, which are occasionally even self-autonomous. This article presents a new epistemological equation, containing a set of new epistemological formulas, that enable future researchers or cyber risk practitioners to conduct epistemological analysis of various cyber risk assessment approaches. The epistemological equation facilitates a pre-designated processes for analysing the effectiveness of a chosen risk assessment approach, analysed through 'Knowledge' and 'Justification' and determining if the chosen approach is designed on 'Truth' or 'Belief', based on the evidence used to formulate the approach (i.e., framework, model, calculator, formula). The contribution from this study—to existing literature on risk assessment, is a new process for uncovering a risk assessment method, based on evidentialism and reliabilism—for risk assessing uncontrollable states in complex IoT systems. This new approach was required to analyse different points of failure in existing cyber risk assessment approaches. The new approach enables risk assessing the risk assessment approaches—which is not commonly considered by engineers when designing various risk assessment approaches. The expected impact of risk assessing existing risk assessment frameworks, models and standards, is a more effective approach for risk assessment of uncontrollable states in complex IoT systems.

The new epistemological equation is enhanced with a comparative empirical analysis that results with a new target state for cyber risk assessment that includes IoT risk based on justification of truth, and a new target state of cyber risk assessment based on the epistemological equation. The target state is supported with a new transformation roadmap for standardisation of IoT risk impact assessment.

The article combines knowledge from common cyber risk assessment approaches and integrates current standards. Hence, the article offers a better understanding of IoT cyber risk, and the interactions in cybersecurity assessment. The findings in this paper constitute;

1. Epistemological analysis of cyber risk assessment approaches for IoT systems;
2. Transformation roadmap for IoT cyber risk assessment; and
3. Dependency describing how IoT companies can achieve their target state.

The roadmap and the design implementation tiers can be applied for:

a. Risk identification, management, estimation, and strategy prioritisation.

In addition, the combination of the outlined steps, presents a process for visualising IoT cyber risk. This was identified as one of the key problems faced by cybersecurity practitioners, because installation of new IoT devices was confirmed as often not reported to the cybersecurity professionals. The visualisation of IoT cyber risk can be used by practitioners and regulators to inform organisations in this space of best practices. The findings are relevant to national and international digital strategies, specifically for IoT cyber risk planning.

### 8.1 Limitations and Further Research

The epistemological equation is based on documented availability. There are additional IoT strategies, cyber risk frameworks, models and methodologies that are not considered in this article—because detailed materials are not publicly available at the time of writing this article.

**Acknowledgements** Eternal gratitude to the Fulbright Visiting Scholar Project.

**Authors Contributions** PR: main author; DDR: supervision; OS, PB: supervision, review and corrections.

**Funding** This work was funded by the EPSRC [Grant Number: EP/S035362/1] and by the Cisco Research Centre [Grant Number CG1525381]. Earlier versions of this work from the combined working papers and project reports prepared for the PETRAS National Centre of Excellence and the Cisco Research Centre can be found in a pre-print online.

**Data Availability** All data and materials included in the article.

**Code Availability** N/A—no code was developed; code was, however, used for running the R Studio analysis.

**Declarations**

**Conflicts of interest** On behalf of all authors, the corresponding author states that there is no conflict or competing interest.

**Publisher's Note** Springer Nature remains neutral with regard to jurisdictional claims in published maps and institutional affiliations.